# Evaluation of the 1077keV γ-ray emission probability from $^{68}$Ga decay

Huang Xiaolong* Jiang Liyang Chen Xiongjun Chen Guochang
China Institute of Atomic Energy, P.O.Box 275(41), Beijing 102413, China

**Abstract:** $^{68}$Ga decays to the excited states of $^{68}$Zn through the electron capture decay mode. New recommended values for the emission probability of 1077keV γ-ray given by the ENSDF and DDEP databases all use data from absolute measurements. In 2011 Jiang Liyang deduced a new value for 1077keV γ-ray emission probability by measuring the $^{69}$Ga(n,2n)$^{68}$Ga reaction cross section. The new value is about 20% lower than values obtained from previous absolute measurements and evaluations. In this paper, the discrepancies among the measurements and evaluations are analyzed carefully and the new values are re-recommended. Our recommended value for the emission probability of 1077keV γ-ray is 2.72±0.16 %.

**KeyWords**: Decay data, γ-ray emission probability, Cross section, Evaluation, $^{68}$Ga

## 1 Introduction

Accurate decay data for $^{68}$Ga are necessary for various applications in nuclear physics and technology. The radionuclide $^{68}$Ga is mainly a positron emitter and is often used in nuclear medicine. It mainly emit annihilation radiation. For its gamma transitions, only the 1077keV γ-ray is important, because the other γ-rays are very weak. The emission probability for 1077keV γ-ray from $^{68}$Ga decay is frequently used to measure the sample activity in activation cross section measurements. Also, GaAs is an important semiconductor with extensive application possibilities in research and industry. Cross section measurements of neutron-induced reactions on GaAs are thus important for characterization of the semiconductor.

The $^{69}$Ga(n,2n)$^{68}$Ga reaction is an important reaction, and has been studied for many years. Most of the experiments use high-resolution γ-ray spectroscopy to count the 1077keV γ-ray from activated samples. Bormann et al. [1], however, have performed a study using a coincidence setup with two NaI(Tl) detectors to count the annihilation γ-rays of the positrons from the $\beta$+ decay of the $^{68}$Ga product nucleus. Among these experiments, a large difference was found between the cross section values which were deduced with annihilation radiation and direct γ-ray spectroscopy respectively(see Fig.1 and 2). Raut's [2] measurements are lower than all other measurements. It is noted that the experimental measurements for the $^{75}$As(n,2n)$^{74}$As reaction show the same behaviour (see Fig.5 in Ref. [2]). Using the ratio of values evaluated with ENDF/B-II to the measured values in $^{75}$As(n,2n)$^{74}$As reaction, the measured values for the $^{69}$Ga(n,2n)$^{68}$Ga reaction were adjusted, as shown in Fig. 2. The adjusted values are in agreement with the values measured using the 1077 keV γ-ray, but still lower than the values measured by Bormann et al. [1].

* e-mail: huang@ciae.ac.cn

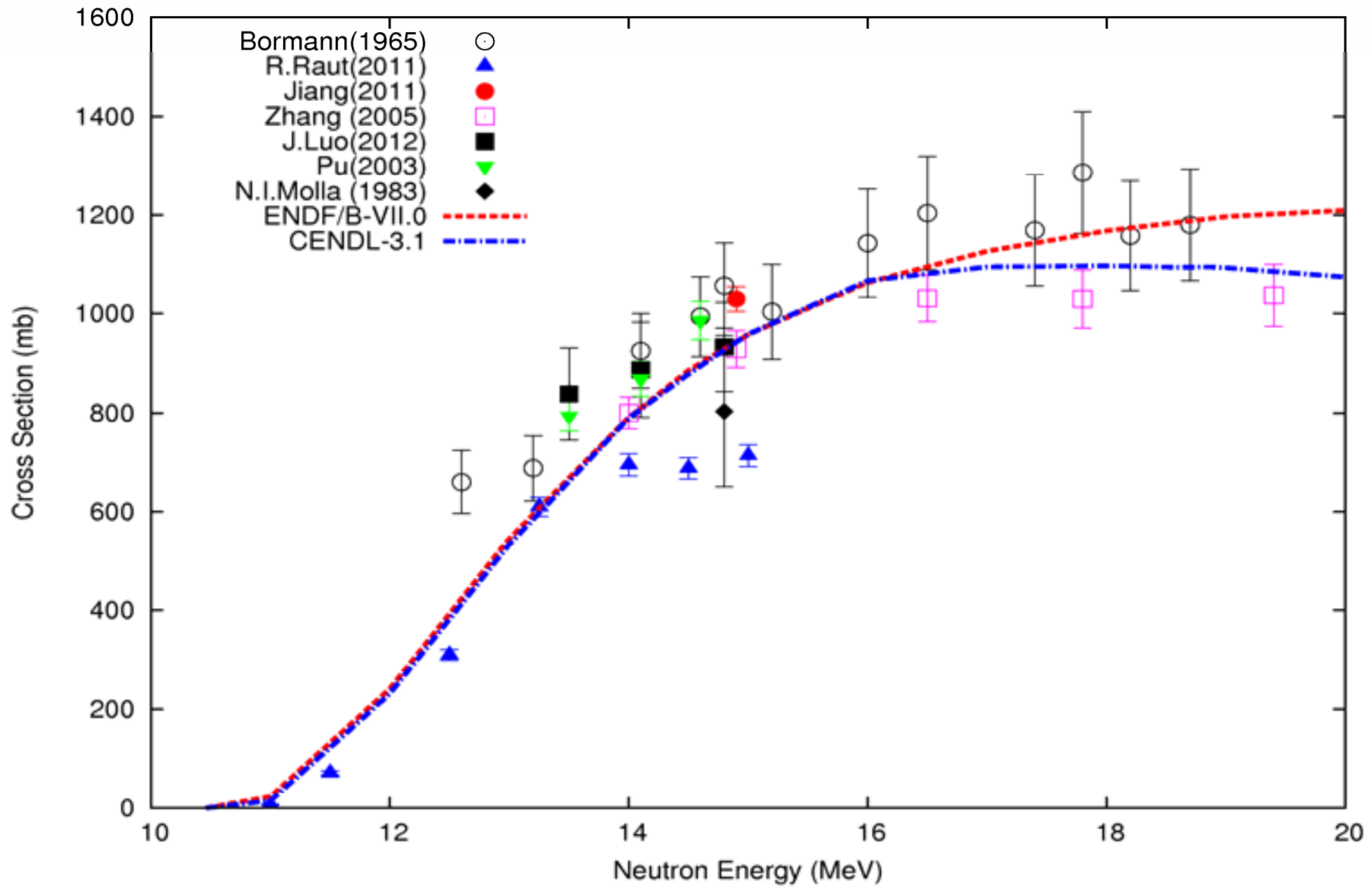

Fig. 1 Measured and evaluated cross section for $^{69}$Ga(n,2n)$^{68}$Ga reaction

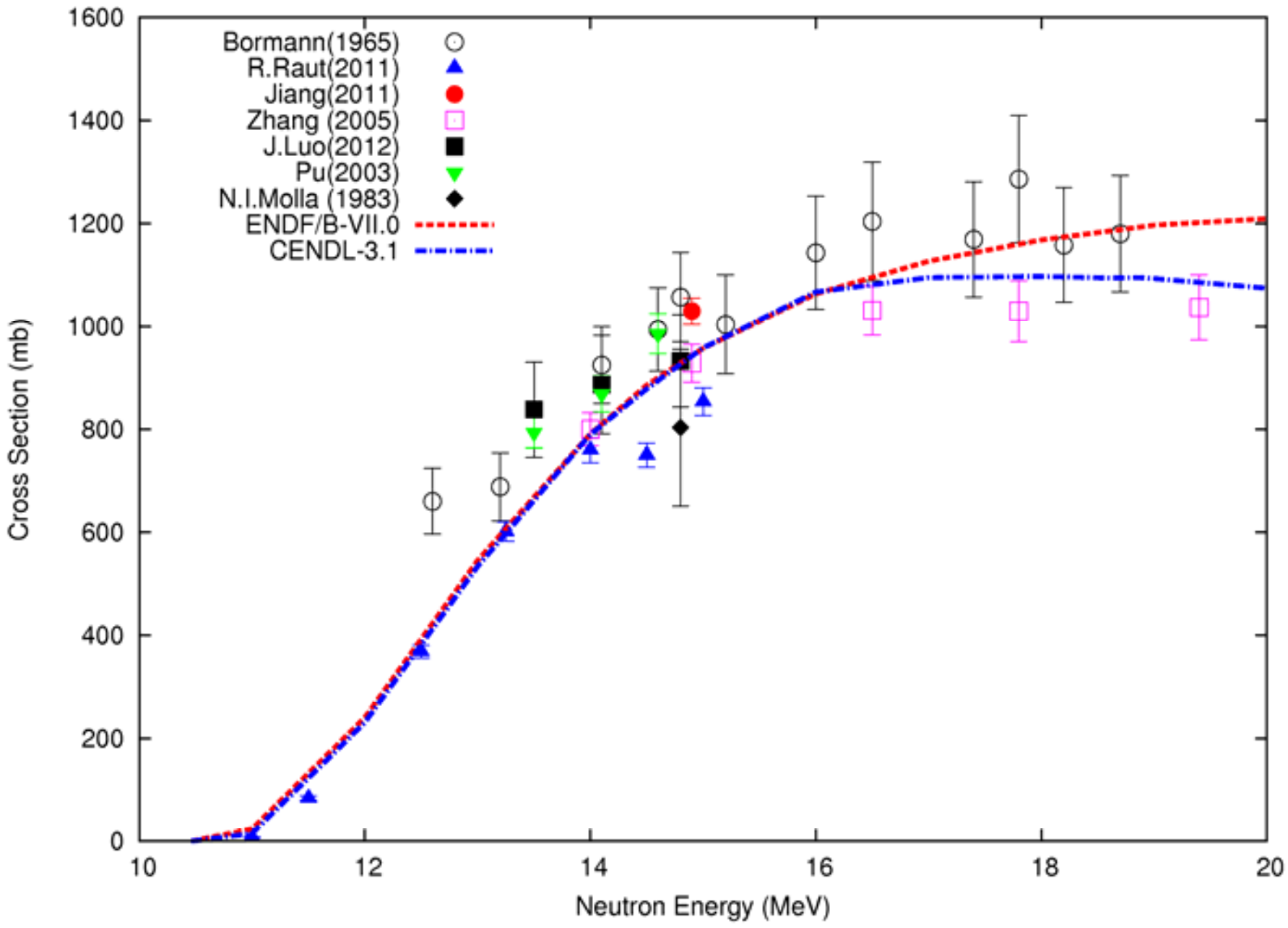

Fig. 2 Adjusted and evaluated cross section for $^{69}Ga(n,2n)^{68}Ga$ reaction

In this work, factors which can induce the difference between these measurements are analyzed. It is found that the most likely reason is due to inadequacies in the current decay data for $^{68}Ga$. Such a difference demands re-evaluation of the 1077 keV γ-ray emission probability.

## 2 Brief introduction to Jiang's measurements of the $^{69}Ga(n,2n)^{68}Ga$ reaction

About 1g of $Ga_2O_3$ powder of natural isotopic composition was pressed to obtain a pellet of diameter 2.0 cm. Monitor foils of Nb (99.9% pure) of the same diameter as the pellets but of weight 0.1 g were then attached to the front and back of each sample. Irradiation of the samples was carried out at the Cockcraft-Walton accelerator in the China Institute of Atomic Energy (CIAE) and lasted 40 minutes. Neutrons were produced by the $T(d,n)^4He$ reaction. The groups of samples were placed at an angle of 0° relative to the beam direction and centered about the T-Ti target at distances of ~3.5 cm. The γ-ray activitiy of $^{68}Ga$ were determined by a high-purity germanium (HPGe) detector.

Cross section of $^{69}Ga(n,2n)^{68}Ga$ reaction at $E_n$=14.9MeV was measured. A large difference was found between the cross section values which were deduced with annihilation radiation and direct γ-rays. Factors which could induce the difference were then analyzed, with the most likely reason being the current decay data for $^{68}Ga$. Using the measured relative counting rates of the characteristic rays, as well as the existing experimental reports, the emission probability for 1077 keV γ-ray in $^{68}Ga$ decay were adjusted to $P_\gamma$ (1077keV)=2.56±0.09 %. The adjusted decay data makes the cross section values deduced with different detection methods agree with each other better. The relevant results are presented in Table 1.

Table 1 Measured cross sections from various decay data for the $^{69}Ga(n,2n)^{68}Ga$ reaction

| $E_n$/MeV | Cross section/mb | $E_\gamma$/keV | **$P_\gamma$ (%)** |
|---|---|---|---|
| 14.9±0.5 | 897±94 | 1077 | **3.0±0.3 [4]** |
| 14.9±0.5 | 1040±27 | 511 | **89.3±1.6 [4]** |
| 14.9±0.5 | 1035±32 | 1077 | **2.56±0.09 [3]** |
| **14.9±0.5** | **1029±25** | **511** | **89.4 [3]** |

The finial result for the $^{69}Ga(n,2n)^{68}Ga$ reaction cross section at $E_n$=14.9MeV was 1030±31(mb) with $P_\gamma$ (1077keV)= 2.56±0.09%.

## 3 Status of experimental data on relative values of γ-ray intensities

In order to clarify the discrepancies in measurements of 1077keV γ-ray emission probability, all the available relative intensities, including Jiang's [3] measurements, are compiled and listed in Table 2, as well as the LRSW results. It is noted that the LRSW results are in good agreement with Jiang's [3] measurements and with the NDS evaluations [12]. This means that the present experimental data on relative value of γ-ray intensities are good.

Table 2 Measured relative γ-ray intensities for $^{68}Ga$

| $E_\gamma$/keV | $I_\gamma$ | | | | | | | |
|---|---|---|---|---|---|---|---|---|
| | Carter[5] | Vaughan[6] | Lange[7] | Vo[8] | Schonfeld[9] | Schotzig[10] | Jiang [3] | LRSW |
| 227 | | | | 0.0037±0.0015 | | | | |
| 483 | | | | 0.0082±0.0009 | | | | |
| 579 | 1.1±0.1 | 0.7±0.1 | 1.00±0.12 | 1.05±0.05 | 1.14±0.15 | | 0.9±0.1 | 1.03±0.04 |
| 683 | | | | 0.0097±0.0006 | | | | |
| 806 | 2.8±0.2 | 2.2±0.2 | 2.95±0.12 | 2.81±0.14 | 2.90±0.31 | 2.97±0.07 | 2.8±0.2 | 2.92±0.05 |
| 939 | | | | 0.0055±0.0005 | | | | |
| 1077 | 100 | 100 | 100 | 100 | 100 | 100 | 100 | 100 |
| 1166 | | | | 0.0005±0.0003 | | | | |
| 1261 | 2.9±0.2 | 3.1±0.2 | 3.00±0.07 | 2.75±0.14 | 3.06±0.31 | 2.91±0.06 | 3.0±0.2 | 2.93±0.04 |
| 1659* | | | | | | | | |
| 1744 | 0.28±0.04 | 0.5±0.01 | 0.30±0.04 | 0.295±0.015 | | | | 0.294±0.013 |
| 1883 | 4.1±0.4 | 4.8±0.3 | 4.33±0.12 | 4.6±0.2 | 3.86±0.59 | 4.14±0.08 | 4.1±0.2 | 4.22±0.06 |
| 2338 | 0.04±0.02 | <0.1 | 0.050±0.006 | 0.031±0.003 | | | | 0.035±0.005 |
| 2821 | | | 0.015±0.002 | 0.0139±0.0011 | | | | 0.0142±0.001 |

*: from Slot [11]; a: (ce+γ)intensity [11].

## 4 Evaluation of the 1077keV γ-ray emission probability

There is only one absolute measurement of the emission probability of 1077keV γ-ray from $^{68}Ga$ ε decay [9]. The $I_{\beta+}$ value was also determined to be 89.14±0.11% in this work. This value is in good agreement with other measured values: $I_{\beta+}$=88±11% [13], $I_{\beta+}$=89.2% [6]. There is little difference in the available measured $I_{\beta+}$ values.

From LOGFT code, the theoretical ratio of $\varepsilon/\beta^+$ can be determined as $\varepsilon/\beta^+$(1077keV level)= 1.505±0.017, and $\varepsilon/\beta^+$(ground state)=0.1017±0.0011. If the measured value $I_{\beta+}$=89.14±0.11% [9] and the relative γ-ray intensities from the LRSW results in Table 2 are used, the emission probability of 1077keV γ-ray can be determined to be $P_\gamma$(1077keV)= 2.81±0.01%. This deduced value is obviously different from the absolute measured value of 3.22±0.03% [9] and Jiang's deduced value of 2.56±0.09%[3].

In order to obtain a reasonable value for the 1077keV γ-ray emission probability, the measured ratio of the total intensity for positrons($p_{ann}$) and for 1077keV γ-ray ($p_{1077}$) in the $^{68}Ga$ ε decay are compiled and listed in table 3. The weighted average results are adopted, as the measured values are inconsistent with each other.

Table 3 Measured and recommended ratio of $p_{ann/}$ $p_{1077}$ for $^{68}Ga$ ε decay

| $P_{ann}/P_{1077}$ | References | Comments |
|---|---|---|
| 54.6±6.0 | Horen [14] | |
| 59.3±6.0 | Carter [6] | |
| 55.3±0.6 | Schonfeld [9] | |
| 69.83±0.38 | Jiang Liyang [3] | |
| 59.8±2.5 | | Unweighted mean |
| 65.6±0.4 | | weighted mean |

| | |
|---|---|
| 62.6±0.7 | LRSW weighted mean |
| 65.6±0.4 | Recommended value, from weighted mean |

From the theoretical ratio of $\varepsilon/\beta^+$(1077keV level)= 1.505±0.017, $\varepsilon/\beta^+$(ground state)= 0.1017±0.0011, the relative γ-ray intensities from the LRSW results in Table 2, and the recommended ratio $P_{ann}/P_{1077}$=65.6±0.4, the 1077keV γ-ray emission probability of can be determined to be $P_\gamma$(1077keV)= 2.72±0.16%. The total emission probability for positrons can also be deduced to be $I_{\beta+}$=89.2±0.2%.

The final value for 1077 keV γ-ray emission probability is recommended to be $P_\gamma$(1077keV)=2.72±0.16 %, and $I_{\beta+}$=89.2±0.2%. These values are obviously different from the absolute measured value of 3.22±0.03% [9] and Jiang's [3] deduced value of 2.56±0.09%.

## 5 Discussion and Conclusions

In order to verify the rationality of our present value for 1077keV γ-ray emission probability, measured cross sections of the $^{69}$Ga(n,2n)$^{68}$Ga reaction around 14.8MeV are adjusted using the present values. The adjusted and original measured reaction cross section are listed in Table 4. It is clear that good agreement is obtained. Therefore, we consider our present value for 1077 keV γ-ray emission probability to be superior to other results, such as the NDS values [12].

Table 4 Comparsion of the original and adjusted measured data for the $^{69}$Ga(n,2n)$^{68}$Ga reaction at 14.8MeV

| References | $E_n$/MeV | Original cross section/mb | **Adjusted cross section/mb** |
|---|---|---|---|
| Molla[15] | 14.8 | 803±153($P_\gamma$(γ1077keV)=3.5%) | **1033±197($P_\gamma$(γ1077keV)=2.72%)** |
| Zhang[16] | 14.9 | 929±37($P_\gamma$(γ1077keV)=3%) | **1025±41($P_\gamma$(γ1077keV)=2.72%)** |
| Luo[17] | 14.8 | 933±90($P_\gamma$(γ1077keV)=3.2%) | **1047±101($P_\gamma$(γ1077keV)=2.72%)** |
| Bormann[1] | 14.8 | 1057±86($I_{\beta+}$=88%) | **1043±85($I_{\beta+}$=89.2%)** |
| **Jiang Liyang[3]** | **14.9** | **1030±31($I_{\beta+}$=89.4%)** | **1032±32($I_{\beta+}$=89.2%)** |

Using our present results, the measured cross sections of the $^{69}$Ga(n,2n)$^{68}$Ga reaction were adjusted. The present recommended decay data makes the cross section values deduced with different detection methods agree with each other better, as shown in Fig. 3.

The excitation function of the $^{69}Ga(n,2n)^{68}Ga$ reaction were then re-evaluated based on the adjusted experimental data. The new values for the $^{69}Ga(n,2n)^{68}Ga$ reaction cross section are in agreement with the adjusted experimental data, but different from the other available evaluations, such as ENDF/B-VII and CENDL-3.1(see Fig. 3).

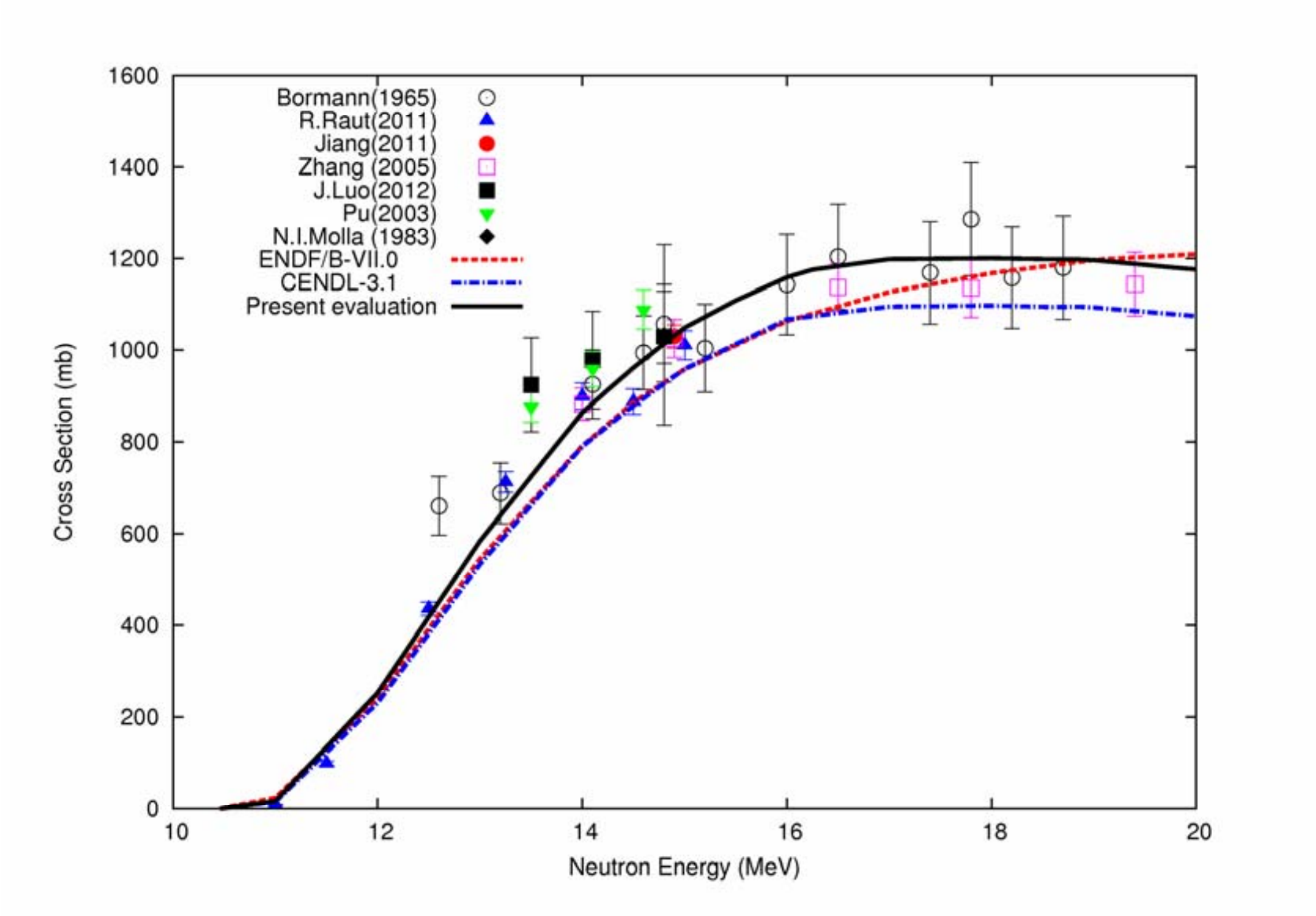

Fig. 3 Adjusted and re-evaluated cross section for the $^{69}Ga(n,2n)^{68}Ga$ reaction

We therefore suggest that the recommended value for emission probability of 1077keV γ-ray from $^{68}Ga$ ε decay should be 2.72±0.16 %. Further work will concentrate on obtaining more precise new measurements to establish the reasonableness of this evaluation.

## References

1 Bormann M, Fretwurst E, Schehka P. Nucl. Phys., 1965, **63**: 438

2 Raut R, Crowell A S, Fallin B. Phys. Rev. C, 2011, **83**: 044621

3 Jiang Liyang. Doctor Thesis of China Institute of Atomic Energy, 2011

4 Lederer C M, Shirley V S. Table of Isotopes, 1978, 7th edition

5 Carter H K, Hamilton J H, Ramayya A V. Phys. Rev., 1968, **174**: 1329

6 Vaughan K, Sher A H, Pate B D. Nucl. Phys. A, 1969, **132** : 561

7 Lange J, Hamilton J H, Little P E. Phys. Rev. C, 1973, **7**: 177

8 Vo D T, Kelley W H, Wohn F K. Phys. Rev. C, 1994, **50**: 1713

9 Schonfeld E, Schotzig U, Gunther E. App. Radiat. Isot, 1994, **45**: 955
10 Schotzig U. App. Radiat. Isot, 1996, **47**: 196
11 Slot W F, Dulfer G H, Molen H. Nucl. Phys. A, 1972, **186**: 28
12 Burrows T W. Nuclear Data Sheets, 2002, **97**: 1
13 Ramaswamy M K. Nuclear Physics, 1959, **10**: 205
14 Horen D J. Phys. Rev., 1959, **113**: 572
15 Molla N I, Islam M M, Rahman M M. INDC(BAN)-002/G, 1983
16 Zhang G H. Private Communication, 2005
17 Luo J, Liu R, Jiang L. Radiochim Acta, 2012, **100**: 231
18 Pu Z, Yang J, Kong X. App. Radiat. Isot., 2003, **58**: 723

## $^{68}$**Ga** 衰变的1077keV γ射线发射几率评价

黄小龙　江历阳　陈雄军　陈国长
中国原子能科学研究院，北京275(41)信箱，102413

**摘要：**$^{68}$Ga衰变通过电子俘获衰变到$^{68}$Zn的激发态。ENSDF和DDEP数据库中新的1077keV γ射线发射几率推荐数据都采用绝对测量数据。2011年江历阳在作博士论文时测量$^{69}$Ga（n,2n）$^{68}$Ga反应截面，推导出1077keV γ射线发射几率比ENSDF和DDEP数据库中的推荐值低20%左右。本文对评价、测量数据间的这种差异进行了细致的分析和新的评价。1077keV γ射线发射几率新的推荐值为2.72±0.16%。

**关键词**：衰变数据，γ射线发射几率，反应截面，$^{68}$Ga